# Modelling Spatial Regimes in Farms Technologies[1]

**Billé AG.[1], Salvioni C.[2] and Benedetti R.[2]**

[1] Faculty of Economics and Management, Free University of Bolzano-Bozen, Bolzano, Italy

[2] Department of Economics, University of Chieti-Pescara, Pescara, Italy

Contact person: Cristina Salvioni salvioni@unich.it

**Abstract** We exploit the information derived from geographical coordinates to endogenously identify spatial regimes in technologies that are the result of a variety of complex, dynamic interactions among site-specific environmental variables and farmer decision making about technology, which are often not observed at the farm level. Controlling for unobserved heterogeneity is a fundamental challenge in empirical research, as failing to do so can produce model misspecification and preclude causal inference. In this article, we adopt a two-step procedure to deal with unobserved spatial heterogeneity, while accounting for spatial dependence in a cross-sectional setting. The first step of the procedure takes explicitly unobserved spatial heterogeneity into account to endogenously identify subsets of farms that follow a similar local production econometric model, i.e. spatial production regimes. The second step consists in the specification of a spatial autoregressive model with autoregressive disturbances and spatial regimes. The method is applied to two regional samples of olive growing farms in Italy. The main finding is that the identification of spatial regimes can help drawing a more detailed picture of the production environment and provide more accurate information to guide extension services and policy makers.

## 1. Introduction

Although theory supports the idea that firms do not operate on the basis of a common production function, i.e., that firms do not use homogeneous technology (Nelson and Winter 1982; Dosi 1988), a global production function is typically proposed in most empirical studies; thus, it is assumed that production technology is invariant over space and across firms. Indeed, especially in land-based industries, such as agriculture, the analysis of production often must be more nuanced and account for the variations in technology arising from locally-specific solutions that satisfy the environmental or social conditions within which firms operate (Mundlak 2001; Just and Pope 2001; Fezzi and Bateman 2011). Under these circumstances estimating a common production function (technology) to all farms can yield biased estimates of the technological characteristics and lead to wrong management advice or policy prescriptions.

For this reason, many empirical studies control for the possibility of heterogeneous technologies by classifying farms into groups on the basis of a priori exogenous information about their technological

---

characteristics, and subsequently estimate separate different production functions for each group. This classification is based on either some a priori information (e.g., location of farms, etc.) or the application of cluster analysis (Alvares *et al*. 2008). It has however been noted (Alvarez *et al*., 2012) that the use of single or even multiple characteristics to split a sample of observations can only be incomplete proxy descriptor for technologies, since differences in technologies can be the result of both observed and unobserved factors.

Modelling unobserved heterogeneity, instead, can be made via non-parametric and flexible mixtures methods. For instance, in several applications to agriculture, finite mixture models provide the opportunity to classify the sample into a finite number of technologies (classes) underlying the data, according to the estimated probabilities of class membership based on multiple specified characteristics or auxiliary/proxy variables (Orea and Kumbakhar 2004; O'Donnell and Griffiths 2006; Alvarez and del Corral 2010; Sauer and Paul 2013; Barath and Ferto, 2015).

The issue of dealing with unobserved heterogeneity has been recently accounted for by an increasing number of methodological papers (Greene 2005a; Greene 2005b; Emvalomatis 2012; Galan *et al*. 2014). Most of them relied on fixed- and random-effects panel data models, which former case capture unobserved heterogeneity through a set of firm specific intercepts that are simultaneously estimated with other parameters. These approaches are increasingly used in the context of stochastic frontier models, though distinguishing between heterogeneity and inefficiency is still an open problem (Amsler and Schmidt 2015, Kumbakhar *et al*. 2014, Colombi *et al*. 2014).

In this article, we focus the attention on a specific type of unobserved heterogeneity, i.e. spatial heterogeneity. Spatial heterogeneity, a term that has been coined by Anselin (1988), can refer to structural instability over space, in terms of changing functional forms or varying parameters (instability in the mean), or to heteroskedasticity, due to different forms of model misspecification that lead to non-constant error variances (instability in the variance)[2]. Spatial heterogeneity can also be classified into discrete heterogeneity and continuous heterogeneity (Anselin 2010). In this article, we are particularly interested in studying discrete spatial heterogeneity, that is situations in which the relationship or the functional form varies across spatial subsets of the data that, in turn, might point to the existence of spatial regimes (Anselin 1988). Spatial regimes are then geographic subsets of data in which the model coefficients assume different values. It can be seen as simply a special case of *group-wise heterogeneity*.

The rationale behind our interest in spatial regimes is that in agriculture there is a long tradition of studies aimed at identifying agro-ecological zones (AEZ) i.e., homogenous and contiguous areas with similar soil, land and climate characteristics and having a specific range of potentials and constraints for land use. The AEZ methodology (FAO 1978) provides maximum potential and agronomically attainable crop yields, under assumed levels of inputs and management conditions. As a matter of fact, within each AEZ the long term, dynamic interactions among site-specific environmental variables which characterize a AEZ and farmer decision making about technology contributed to develop local specific varieties and production technologies, hence to give rise to local technology clusters, a concept similar to that of *terroir* used in oenology. In other words, environmental and social factors contribute to describe a patchy technological landscape. Borrowing from spatial econometric analysis, these local technology clusters can be defined as spatial regimes in farms technologies. However, the zoning of such local technology clusters or terroirs is largely unknown. This is mainly due to the fact that their identification needs comprehensive spatial modelling of soil, agronomical and climatic properties, including their changes through time, hence the processing of large quantities of data acquired at a very fine spatial resolution. In fact, researchers can often rely on a few control variables. For example, agricultural surveys usually do not carry information about cultivars grown, which are key to correctly associate the climatic variables collected in local meteorological stations with farms, in consideration of the differences in the timing of the phenological stages of each cultivar.

---

[2] Note that in this paper we only consider heterogeneity due to structural instability in the mean.

These issues motivate us to endogenously identify spatial regimes in farm technologies. To that end, the idea is to exploit the information derived from geographical coordinates (longitude and latitude) to approximate the effects of a variety of complex, dynamic interactions among site-specific environmental variables and farmer decision making about technology that are often not observed at the farm level. The starting point of our analysis is the hypothesis that there may be significant spatial dependence (Anselin 2002) and spatial heterogeneity in the natural clustering of farms around different geographical poles of attraction. With the aim of accounting for both spatial dependence and spatial heterogeneity in the estimation of the production function of olive-growing farms in Italy, in this article we apply an approach based on the recent works by Andreano *et al.* (2016) and Billé *et al.* (2017).

A main difficulty with cross-sectional data, as stressed by Anselin (2010), is that it is often hard separating spatial heterogeneity from spatial dependence. Spatial dependence is viewed as a special case of cross-sectional dependence for which a parametric structure of the covariance matrix is imposed by using a specific ordering of the spatial units. The essence of the problem is that cross-sectional data, while allowing the identification of clusters and patterns, do not provide sufficient information to identify the processes that led to the patterns. As a result, it is impossible to distinguish between the case where the cluster is derived from an apparent contagion due to structural change, hence discrete heterogeneity, or follows from a true contagious process, hence dependence. Moreover, when we deal with unobserved spatial regimes, no a priori information is typically available to identify them and to easily estimate the econometric model.

In this study, we employ a two-step estimation approach. In the first step, we employ the underlying spatial heterogeneity as the criterion to divide the whole sample into groups of contiguous observations that are homogeneous in terms of technology. We capture potential coefficient variations in the the estimated function by making use of an algorithm that builds on local estimation procedure, in the spirit of Cleveland and Devlin (1988), in conjunction with the adaptive weights smoothing (AWS) approach, see Polzehl and Spokoiny (2000). The AWS procedure, originally proposed in the context of image denoising, can be used in all applications where the regression function is likely to possess jumps or sharp edges. We use the AWS approach to make the local coefficients statistically converge into a discrete variation over space, hence identify the edges between different local technology clusters. In the second step of the estimation strategy, once the spatial regimes in technologies are identified, we contemporaneously account both for spatial regimes and spatial dependence (which in this case we suppose is determined by a true contagion process) with the aim to reduce as much as possible the consequences of model misspecifications. In the current research, we employ the above two step approach to investigate farm-level olive production in two regions in Italy, i.e. Tuscany and Apulia.

The remainder of the article is organized as follows. In section 2 we explain the iterative procedure used to detect unobserved spatial heterogeneity and to identify the spatial technological regimes in a data-driven approach, we then specify the spatial econometric model with regimes used to simultaneously detect the effects of spatial dependence and spatial regimes, and finally we explain how we address the potential endogeneity issue that typically arises into production function estimation. In section 3 we introduce the data set and show the main estimation results of the application of the procedure to two regional samples of olive growing farms in Italy. Finally, section 4 concludes.

## 2. The Model

In this section, we present the procedure used to detect unobserved spatial heterogeneity and to identify groups of spatial units (farms in our case) with homogeneous local production function parameters, i.e. spatial regimes in technologies. In particular, in subsection 2.1 we briefly explain the iterative procedure which allows us to control for unobserved spatial heterogeneity and to identify spatial regimes in a data-driven approach. In subsection 2.2 we specify the spatial econometric model with regimes used to simultaneously detect the effects of spatial dependence and spatial regimes.

Whereas in subsection 2.3, we present the production function that we are going to estimate and explain how we address the potential endogeneity issue.

## 2.1 The iterative procedure used to identify spatial regimes

The iterative procedure used to identify spatial regimes is based on a first set of locally (geographically) weighted estimates (Cleveland and Devlin 1988), and the adaptive weights smoothing (AWS) approach (see Polzehl and Spokoiny 2000). The result of the combination of above-mentioned methods is an algorithm that is able to endogenously identify (i.e., in a data-driven approach) regimes over space, that is large homogeneous areas with sharp discontinuities. Overall, the iterative procedure tries to answer the following question: do the coefficient estimates of a spatial unit (farm in our case) statistically differ from the coefficient estimates of the nearby ones? Details on the proposed algorithm can be found in Andreano *et al.* (2017). Further improvements as well as the R code can be found in Billé *et al.* (2017). In this article, we adopt the same two-step procedure as in Billé *et al.* (2017). A brief explanation of the iterative procedure follows.

To account for local parameter estimates that vary over space, simple linear functions may provide a reasonable approximation of the local estimate as long as we use the information on a group of observations (farms in our case) close to observation $i$. The goal is to approximate the following model

$$y = (\beta \odot X)\mathbb{1} + \varepsilon, \varepsilon \sim iidN(0, \sigma_\varepsilon^2 I) \tag{2.1}$$

where $y$ is an $n$-dimensional column of dependent variables, $\beta$ is an $n$ by $k+1$ matrix of local coefficients with i-th row $\beta_i = (\beta_{i0}, \beta_{i1}, \beta_{i2}, \dots, \beta_{ik})'$ referred to observation $i$, $\odot$ is the Hadamard product operator in which each element of $\beta_i$ is multiplied by the corresponding element of the i-th row in the matrix of regressors $X$, $x_i = (x_{i0}, x_{i1}, x_{i2}, \dots, x_{ik})$, $\mathbb{1}$ is a $(k+1)$-dimentional column vector of ones, and $\varepsilon$ is an $n$-dimensional column vector of *iid* normal innovations.

Local estimates of $\beta_i$ (i.e. for each unit in space) can be directly obtained by repeated weighted least squares (WLS)

$$\hat{\beta}_i^{WLS} = (X'W_i X)^{-1} X'W_i y, \quad i = 1, \dots n \tag{2.2}$$

where $W_i$ is an $n$-by-$n$ matrix whose non-zero elements denote the geographical distance weights $(w_{i1}, w_{i2}, \dots, w_{ij}, \dots, w_{in})$, obtained on the basis of longitude and latitude, of each of the $n$ observed data for observation. So, the role of the $W_i, i = 1, \dots, n$, weighting matrices is simply to give different distance-based weights in estimating the local coefficients *i*. We therefore have *n* diagonal spatial weighting matrices, one for each of the *n* observed data. The final result of the locally weighted least squares is obtained when *n* sets of local parameter estimates, that correspond to the local marginal effects, have been estimated. The subsets of observations (farms in our case) i.e., the neighborhood of *i* used for each local weighted least square fit, are selected by the "bandwidth" or "smoothing parameter" that determines how much of the data is used to fit each local regression.

Let now define $y^{(i)} = X\beta_i + \varepsilon^{(i)}$, $\varepsilon^{(i)} \sim N(0, \sigma^2_{\varepsilon^{(i)}} I^{(i)})$, the model effectively and iteratively estimated for each $i = 1, \ldots, n$. The WLS estimator for each observation *i* in equation (2.2) has the following expected value and variance-covariance matrix

$$E(\hat{\beta}_i) = E[(X'W_i X)^{-1} X'W_i y^{(i)}] = E[(X'W_i X)^{-1} X'W_i (X\beta_i + \varepsilon)] = \beta_i$$

$$\begin{aligned} V(\hat{\beta}_i) &= E\left[(\hat{\beta}_i - \beta_i)(\hat{\beta}_i - \beta_i)'\right] = E\left[((X'W_i X)^{-1} X'W_i y^{(i)} - \beta_i)((X'W_i X)^{-1} X'W_i y^{(i)} - \beta_i)'\right] \\ &= E[((X'W_i X)^{-1} X'W_i (X\beta_i + \varepsilon) - \beta_i)((X'W_i X)^{-1} X'W_i (X\beta_i + \varepsilon) - \beta_i)'] \\ &= E[((X'W_i X)^{-1} X'W_i \varepsilon)((X'W_i X)^{-1} X'W_i \varepsilon)'] \\ &= (X'W_i X)^{-1} X'W_i E(\varepsilon\varepsilon') W_i X (X'W_i X)^{-1} \end{aligned}$$

so that the WLS estimator is locally unbiased (i.e. given the optimal bandwidth value) with variance-covariance matrix in equation $V(\hat{\beta}_i) = \sigma^2_{\varepsilon^{(i)}} (X'W_i X)^{-1} X'W_i^2 X (X'W_i X)^{-1}$.

Given our aim to identify spatial regimes i.e., large homogeneous regions separated by sharp discontinuities, we use the AWS approach to detect the greatest possible local neighborhood of every observation *i* in which the local parametric assumption is justified by the data. The AWS approach is based on a successive increase of local neighborhoods around every point *i* and a description of the local model within such neighborhoods by assigning weights. The weights describing the shape of the local model at the point *i* depend on the result of the previous step of the procedure. More in detail, we iteratively update the weights (geographical distances) in the main diagonal of $W_i$ and, at each iteration, we compare the estimated beta coefficients in (2.2) by using Wald test statistics in order to check whether pairs of spatial units (farms in our case) follow the same economic behavior (production function in our case). The initial $W_i$ matrix, i.e., the weights from which we start the procedure, is defined by using a bi-square kernel weighting function, whereas a Gaussian kernel weighting function is used during the updating procedure of the weights. Both these kernels are defined as functions of the Euclidean distance[3] between observations i and j, i.e. $d_{ij}$, and the bandwidth *b,* though alterative definitions of distances can be used.[4] In this article, we use an *adaptive bandwidth* because it ensures sufficient (and constant) local information for each local calibration[5]. We select the optimal adaptive bandwidth value ($b^{opt}$) by a mix selection of the model specification for local estimates, the optimizing criterion (AIC in our case; see Fotheringham, *et al.* 2002, page 61) and the used kernel which, consistently with the function chosen to define the initial weights, we choose to be a bi-square kernel function. To this purpose, we used the package *GWmodel* in R (Lu et al. 2014). Details on the steps of the iteration procedure are in the appendix.

---

[3] Alternative definitions of distances, like e.g. economic-based distances, can be defined.

This procedure does not impose any restrictions and is fully adaptive in the sense that no prior economic information about the spatial structure is required.

## 2.2 Controlling for spatial dependence

Once the spatial regimes are identified by applying the iterative procedure described above, i.e., after controlling for spatial heterogeneity, we have to deal with spatial dependence among the observations. A huge literature is available on several types of spatial autoregressive models that can be alternatively used to control for spatial interactions (see e.g. LeSage and Pace 2009). Among the most general ones, the spatial autoregressive model with autoregressive disturbances (typically known as SARAR/SAC model) is one possible candidate, which includes both the spatially lagged dependent variables and a parametric correlation structure among the disturbances. Asymptotic properties of the maximum likelihood and quasi-maximum likelihood estimators has been proved by Lee (2004) and of the generalized method of moments by Kelejian and Prucha (2010). By adding the information about spatial regimes, this model can be written as

$$\begin{bmatrix} y_1 \\ \vdots \\ y_c \end{bmatrix} = \rho W_1 \begin{bmatrix} y_1 \\ \vdots \\ y_c \end{bmatrix} + \begin{bmatrix} X_1 & \cdots & 0 \\ \vdots & \ddots & \vdots \\ 0 & \cdots & X_c \end{bmatrix} \begin{bmatrix} \beta_1 \\ \vdots \\ \beta_c \end{bmatrix} + \begin{bmatrix} u_1 \\ \vdots \\ u_c \end{bmatrix}, \begin{bmatrix} u_1 \\ \vdots \\ u_c \end{bmatrix} = \lambda W_2 \begin{bmatrix} u_1 \\ \vdots \\ u_c \end{bmatrix} + \begin{bmatrix} \varepsilon_1 \\ \vdots \\ \varepsilon_c \end{bmatrix} \quad (2.3)$$

where $\tilde{y} = \{\tilde{y}_\square\}$ is a partitioned $n$-dimensional column vector of dependent variables (production in our case), $\tilde{y}_j = \begin{bmatrix} y_j \\ \vdots \\ y_{n_j} \end{bmatrix}$ for $j = 1, \dots, c$ is the $n_j$-dimensional vector of dependent variables relative to the unknown regime (cluster) $j$, with $n = n_1 + \cdots + n_j + \cdots + n_c$; $\rho$ and $\lambda$ are the well-known spatial autoregressive coefficients, which capture the spillover effects in the dependent variables and the correlation structure among the disturbances, respectively; $W_1$ and $W_2$ are $n$ by $n$ spatial weighting matrices; which pre-specify the spatial structure among the observations; $\tilde{\beta} = \{\tilde{\beta}_j\}$ is the $(k \times c)$-dimensional partitioned column vector of parameters (a vector for each cluster); $\tilde{X} = \{\tilde{X}_j\}$ is a block-diagonal matrix of $k$ regressors (inputs in our case) of dimension $n$ by $(k \times c)$; $\tilde{\varepsilon} = \{\tilde{\varepsilon}_j\}$ is a partitioned $n$-dimensional column vector of *iid* normal innovations, with $\tilde{\varepsilon}_j \sim N\left(0, \sigma_{\tilde{\varepsilon}_j}^2 I_{n_j}\right)$ each. Note that the weighting matrices are not partitioned, so that we do not assume unknown form of *group-wise autocorrelation heterogeneity*, but rather observations that belong to different clusters may be connected each other. The grade of connection directly depends on the choice of the criteria used for defining the weights. If dense matrices are assumed, all the $n$ observations are connected with a different weight. Model identification issues require either $W_1$ and $W_2$ to be different or the vector $\tilde{\beta}$ statistically significant.

---

[4] The choice of the value of b is crucial because it determines, first, which observations receive weight in the local estimate and, second, how rapidly the weights decline with distance. In other words, b defines the search window size. In general, higher values of b put more weight on distant observations, leading to results similar to those obtained by OLS, which are more biased for the local estimation procedure but also more efficient. On the contrary, if the bandwidth value tends to zero, then the local estimation is performed by using a smaller number of observations which reduces the local bias but increases the standard error.

[5] It is worth nothing that when b is fixed all the observations within the bandwidth are taken into account for the local neighborhood definition. When b is a variable (adaptive) bandwidth, a fixed number of neighboring observations is assumed, thus defining a k-nearest neighbor approach.

The model in equation (2.3) reduces to a spatial autoregressive error (SAE) model with regimes by setting $\rho = 0$ and to a spatial autoregressive (SAR) model with regimes by setting $\lambda = 0$. Both the squared weighting matrices, i.e. $W_1$ and $W_2$, have diagonal elements $w_{ii}$ equal to zero, i.e. each spatial unit is not viewed as its own neighbor, and they are normalized so that the admissible parameter spaces of $\rho$ and $\lambda$ are known and less than unity in absolute value. In this article, we adopt the typical row-normalization rule for the weighting matrices.

It is worth nothing that robust covariance estimates can be properly obtained by models with heteroskedastic and autocorrelated consistent (HAC) structures (see Kelejian and Prucha 2007 for a spatial extension), which are able to properly handle unobserved factors from an econometric perspective (Dell *et al.* 2014; Deschênes and Greenstone 2007).

**2.3 The estimation of a production function and potential endogeneity**

In this article, we apply the iterative procedure described in sections 2.1 to identify the unobserved spatial production regimes, i.e. spatial groups of farms that are homogeneous in terms of technologies. For this purpose, we replace the general model in 2.1 with the following Cobb-Douglas production function in logs

$$y = \beta_1 x_1 + \beta_2 x_2 + \beta_3 x_3 + \beta_4 x_4 + \varepsilon \qquad (2.4)$$

where $y$ is a column vector of logs of produced quantities and and $x_1, x_2, x_3, x_4$ are column vectors of logs of area (land used), capital inputs, labor inputs and intermediate inputs, while $\varepsilon \sim N(0, \sigma_\varepsilon^2 I_n)$. Once the spatial regimes are identified, we also control for spatial dependence by estimating the spatial model with regimes (2.3). In the context of production function models the *endogenous spatial interactions* i.e., $\rho$ capture the spatial dependence structure between farms' production systems (Horrace *et al.* 2016). For example, farmers' decisions on production innovations that spill-over among neighboring farms. The coefficient $\lambda$ of the spatially autoregressive disturbances may capture an *unobserved dependence structure*, so that a shock on a farm production propagates through the entire system of farms' production with a higher effect on the neighborhood.

When we consider the production function, a typical problem of endogeneity arises for one or more of the regressors (Ackerberg *et al.* 2015; Shee and Stefanou 2014; Latruffe *et al.* 2016). The endogeneity of inputs is mainly due to the presence of some determinants of the firm productions that are unobserved to the econometrician (omitted-variable bias) but observed by the farmer, leading to inconsistency of the standard least squares estimators. In a linear framework, the standard approach for addressing the potential endogeneity bias is to use instrumental variables. The two-stage least squares (2SLS) estimator (Hansen, 1982) has been widely used for this purpose. To efficiently combine 2SLS with repeated local estimates, we regress the set of endogenous variables in (2.4) on the selected instruments *before starting* the iterative procedure in order to guarantee new orthogonality conditions. To briefly explain, consider the previous model of interest (2.4) in a generic form, $y = X_1\beta_1 + X_2\beta_2 + \varepsilon$, where $X_1$ are $k_1$ exogenous variables, whereas $X_2$ are $k_2$ endogenous variables. Given some useful information on a set of $k_3$ instruments for $X_2$ — say $Z = [X_1, P]$ with $k_P \geq k_2$ and $P$ other instruments different from the exogenous ones in the model — we can obtain the predicted values of $X = [X_1, X_2]$ from the first step as $\hat{X} = Z(Z'Z)^{-1}Z'X$. These values are then used in the local regressions during the iterative procedure with $\hat{\beta}_i^{WLS} = (\hat{X}'W_i\hat{X})^{-1}\hat{X}'W_i y$.

Controlling for endogeneity becomes more difficult if we wish to concurrently consider spatial spillover effects (Rey and Boarnet 2004) due to the presence of two different sources of endogeneity, namely, endogenous regressors and spatially lagged variables. The endogeneity problem within the spatial framework can be addressed, for example, by using the spatial heteroscedasticity and autocorrelation consistent (SHAC) estimator proposed by Kelejian and Prucha (2007). In fact, in our local estimates endogeneity of the regressors is an issue prior to the application of the convergence procedure that allows us to endogenously identify the spatial regimes. During the first stage of the estimation procedure we only control for heterogeneous effects. It is only after this first stage that we estimate both a spatial error model and a spatial error model with regimes. Under these circumstances, in the first-stage estimates, we can appropriately address endogeneity by using the generalized instrumental variable method without incurring in the problems deriving from the concurrent existence of another source of endogeneity.

## 3. An application to olive farms in Italy

We use the above technique to examine the production of olives in Italy. Italy ranks second in the world after Spain for olive oil production. The Italian olive sector is still characterized by a large number of small operations. Specifically, Italy has the highest number of holdings (776 000) with the smallest average size (1.3 ha) in the European Union (EU) Mediterranean countries. Over time, the different microclimate conditions, soil formations and elevation levels have led to natural or manmade modifications (breeding and selection) of the olive tree into many location-specific varieties, each with different productivity levels, agronomic needs and adaptability to irrigation and mechanization. In the case of arable crops (e.g. maize and oil seeds), local varieties have been widely substituted by industrial global varieties whose production response does not vary greatly over space. In the case of olive trees, locally-specific varieties are still largely in use. The territorial anchorage of the production of location-specific olive varieties is further strengthened by social and marketing considerations because, similarly to the case of wine and grapes, farmers choose varieties on the basis of not only agronomic characteristics (e.g., disease resistance, climate preferences, high productivity) but also the aptitude for preserving local production knowledge (e.g., flavor, suitability for curing, etc.) and ability to guarantee the production of high-quality oil.

Based on all these considerations, it follows that the technology available to farms depends on the characteristics of the physical, social and economic environment in which production takes place. In other words, the underlying production technology is not the same for all olive farms; rather, it is location specific, and the group of farms sharing the same technology can be termed a local technology cluster, a concept similar to that of *terroir* in oenology. The local olive production function used by farmers operating in a territory results from the choice of locally-optimal technology from a given menu of technologies as a consequence of a process of localized technological change (Stiglitz and Atkinson 1969; Nelson and Winter 1982; Antonelli 2008; Acemoglu 2015). This view is consistent with evolutionary theories (Nelson and Winter 1982; Dosi 1988) according to which firms cannot be assumed to operate using a single common production function. These theories explain why permanent asymmetries exist across firms in terms of production technologies and quality of products (Dosi 1988). External inputs and the past accumulation of skills and knowledge guide the creation of technological knowledge. The technology prevailing in the local technology cluster is the efficient solution to the specific techno-economic problems experienced by the firms operating in the cluster. This solution consists of specific families of recipes and routines, and it is based on carefully selected principles derived from natural sciences jointly with specific rules aimed at acquiring related new knowledge (Dosi and Nelson 2013).

For example, in the case of olive trees, farmers grow different cultivars that have different yields, aptitude to the mechanization of the harvesting and input needs. As we noted above, the decision to grow a low-yield or higher-cost variety is partly connected to the characteristics of the local natural environment (e.g., climate, water, soil) and partly to the preservation of the local cultural heritage (e.g., flavor and quality of oil, landscape). Under these circumstances, it is difficult, if not impossible, to collect all the information needed to define the boundaries of the local technology cluster. A solution is offered by the application of the two-step procedure described in section 2.

**3.1 Data**

This study relies on cross-sectional data collected by the 2013 Italian Farm Accountancy Data Network (FADN) survey. The data set consists in 233 observations for Tuscany and 227 for Apulia. The FADN sample is stratified according to the criteria of the geographical region, economic size and type of farming. The field of observation is the total number of commercial farms, that is farms large enough to provide a main activity for the farmer and a level of income sufficient to support his or her family. The survey gathers physical and structural data (e.g., location, crop areas, livestock numbers, labor force) and the economic and financial data needed for the determination of incomes and business analysis of agricultural holdings. We also exploit a peculiar characteristic of the Italian FADN, which provides longitude and latitude coordinates of farms, thus allowing us to use spatial econometric models and, in our specific case, account for farm spatial heterogeneity. Another advantage of the Italian FADN is that the information on input use and production results is collected by activity. Consequently, unlike most previous studies (see for example Dinar *et al*., 2007 and Karagiannis and Tzouvelekas, 2009 for applications to olive production), we do not need to rely on data that refer to the entire farm production. Instead, both output results and input use in our study are activity specific. The availability of information by activity has several advantages. First, we do not need to focus on farms that are highly specialized in the production of a specific crop — olives, in our case — thus preventing the loss of observations related to farms that grow the selected crops jointly with other products and allowing us to segment the regional samples in groups of farms large enough to produce reliable estimates. Second, given that we do not have to aggregate the farm crop outputs, we can measure output in physical rather than in monetary terms. This feature, along with the availability of activity-specific input uses, guarantees more accurate results.

The dependent variable in the production function in equation (2.4) is the olive production measured in kilograms. The inputs included as explanatory variables are (a) *land* (A) measured in hectares, including only the share of utilized agricultural area devoted to olive tree cultivation; (b) *labor* (L), comprising hired (permanent and casual) and family labor, measured in working hours; (c) *capital* (K), proxied by the hours of mechanical work employed in olive growing and harvesting; and (d) *intermediate inputs* (M) — including expenses for water, fertilizers, pesticides, fuel and electric power and other miscellaneous expenses — measured in euros and augmented with the expenses for contract work. The descriptive statistics are reported in Table 1.

INSERT TAB. 1 HERE

To control for the potential endogeneity of inputs, we tested the suitability of a set of variables to be used as instruments. All the inputs, i.e. labor, capital, and intermediate inputs, are treated as endogenous. Recall that we proxy capital by the hours of mechanical work employed in olive growing, so we have to treat it as an endogenous variable. This set first contains, following the standard approach proposed by Pindyck and Rotenberg (1983), the lagged values of the three variables suspected of being endogenous, namely $L_{t-1}$, $K_{t-1}$, and $M_{t-1}$. Additionally, we consider the prices of nutrients (PF) and pesticides (PP)[6], the opportunity cost of labor[7] (OCL) and capital[8] (OCK) provided in the Italian FADN, and a proxy of the investments in machinery (I) that is obtained as the variation in horsepower at the farm level observed between 2013 and 2012. After testing for the correlation among the endogenous variables and the above-mentioned available instruments by using simple linear regression models, we find that in the cases of Tuscany and Marche, the lagged values of inputs ($L_{t-1}$, $K_{t-1}$, and $M_{t-1}$) are valid instruments of the three variables suspected of being endogenous, while in Apulia, the valid instruments are the lagged value and opportunity cost in the case of labor ($L_{t-1}$, OCL); lagged value, opportunity cost and investment in machinery for capital ($K_{t-1}$, OCK, I); and finally, lagged value and price of fertilizers ($M_{t-1}$, PF) in the case of intermediate inputs.

**3.2 Results**

Our empirical results are based on year 2013, while the 2012 data set is used to check the robustness of the grouping of farms. To model production, we first correct for endogeneity, then we apply the iterative procedure described in section 2.1 by using the Cobb–Douglas (CD) functional form in equation (2.4) in order to identify the spatial production regimes by repeated WLS. Finally, we estimate the model specified in (2.3) and its simplified version without regimes, to control for possible spatial dependence not already controlled for. We also estimate the CD model by OLS, with and without regimes, as benchmark models to check if including spatial effects improves the goodness of fit.

---

[6] The exogenous property of input prices can be reasonable assumed since farms in our sample are small-size enterprises, as a consequence they do not have market power in input markets.
[7] Wage per hour in agriculture defined by collective negotiations at the provincial level.
[8] Cost of hiring machines operated by the farm's labor.

The optimal bandwidth values, i.e., the optimal numbers of nearest neighbors required for local estimates, suggested by the AIC and used to fit the models with regimes are 67 for Tuscany and 111 for Apulia. These values come from a proper combination of the assumed economic model specification and the trade-off problem between bias and variance in considering local estimation procedures, given the sample size. Therefore, the optimizing bandwidth criterion (AIC in our case) tries to find the best subsample size that is able to give the best local fit of the specified model, see equation 2.4, for each point regression $i$. It is worth noting that the model in equation 2.4 should be correctly specified.

We identify 4 spatial regimes in Tuscany (Figure 1a) and 3 in Apulia (Figure 1c).

INSERT FIG. 1 HERE

The results show that the relationship between input use and olive output does not vary evenly across space; rather, it describes a patchy landscape due to spatial heterogeneity. After all, it is not surprising to find, for example, clusters of lower output elasticity of mechanical work on steeply sloping land or higher output elasticity of land when tall olive trees, i.e., those not suitable for mechanical harvesting and pruning, are grown. Input use and yields may also be locally influenced by private and public regulations. For example, farms located in areas covered by geographical indication such as Protected Denomination of Origin (PDO) may be required to comply with the requirement to limit yields to improve the quality of certified products, while those operating in environmentally sensitive areas may be required to reduce their nutrient loadings. At this regard, it is interesting to note that in both regions in both regions the spatial regimes conform well with expert based maps of the olive oil PDO. PDO rules — which aim to protect the names of quality agricultural products — define the geographic area covered and regulate the varieties of olives (usually local cultivar) that can be grown, and the methods of cultivation that can be used. As we already mentioned the local cultivars are the result of a long run process of selection and adaptation to the environment. The high degree of overlap between the clusters identified by the iterative procedure and the geographic distribution of local cultivars and PDO cross-validates our results and the suitability of geographic coordinates as a proxy for variety variations, soil type, climate, etc. In other words, this overlap validates the ability of farms' longitude and latitude to account for the heterogeneity in spatial distribution of technology that, in turn, arises from the adaptation of production techniques to variables that are not usually detected by agricultural surveys at the farm level. In other word, environmental and social factors may contribute to describe a patchy technological landscape characterized by the presence of discrete spatial technological regimes.

To statistically assess the significance of the groupings of farms identified by the iterative procedure, i.e., to test whether the beta coefficients statistically differ across clusters, we computed the Chow test and its spatial version (Anselin 2005). In Tuscany, the results (Table 2) indicate strong evidence of significantly different coefficients in each of the spatial clusters of farms and suggest that the proposed partition can be accepted. In Apulia, the AIC criterion signals that the global model fits the data best. It is worth noting, however, that the (admittedly slightly) significant spatial Chow test statistic (p-value: 0.079) suggests that the clusters of farms identified by the iterative procedure can still be used for descriptive purposes and further investigation.

INSERT TAB. 2 HERE

Finally, in order to check the robustness of the grouping of farms to short term variations in the output and inputs variables, due for example to variations in weather conditions, we tested the sensitivity of the clustering to the change in time by replicating the application of the iterative procedure on two regional samples of olive growing farms referred to year 2012 (Figure 1b and 1d). It is easy to see that the zoning obtained using the data referred to year 2013 is very similar to the one obtained using the sample referred to the previous year. The minor differences in the borders are largely due to the fact that the samples are not balanced, hence part of the farms observed in year 2012 were replaced by different farms with different locations. In other words, as expected, we find the clustering of farms is time invariant, the reason for this is that it is the result of the long-term interactions of social and environmental variables that affect the farmers' choice of technology.

After having controlled for spatial heterogeneity and identified the local technological clusters of farms, we estimate the model specified in (2.3) to control for possible spatial dependence in the data not already controlled for during the application of the iterative procedure. We did not find any statistically significant spatial autoregressive coefficient ($\rho$) in our empirical application, which result was also supported by the LR test statistic values in Table 6. As already noted, when $\rho = 0$, then the SARAR/SAC model described in (2.3) collapses to a spatial autoregressive error (SAE) model. The coefficient estimates of the global SAE and those of the SAE with regimes are reported in Tables 3-4. At this regard, it is worth noting that, like in OLS, the coefficient estimates of both SAE and a SAE with regimes can be interpreted as output elasticity. This is because the autocorrelation among disturbances does not affect the expected mean, but only the variance-covariance matrix of the spatial model, hence we do not need to calculate the direct, indirect and total effects typically needed for spatial marginal effects interpretations (LeSage and Pace 2009).

INSERT TAB. 3 HERE

INSERT TAB. 4 HERE

The best fitting model in each region has been defined on the basis of the smallest AIC (table 5). The best-fitting model in Tuscany is the one accounting for both spatial dependence in the disturbances ($\lambda$) and spatial heterogeneity. In fact, it is notable that in Tuscany, the measure of spatial error dependence is highly statistically significant both in the global model and the model with regimes though slightly lower in magnitude in the latter model. This means that spatial dependence is still at work even after controlling for spatial heterogeneity. Also in Apulia, we find spatial error dependence is statistically significant, though in this case the AIC criterion selects the global spatial model as the specification that fits the data best. This result does not mean that spatial heterogeneity is not at work in Apulia, it rather indicates that the model with spatially varying parameters is excessively complex, i.e. less parsimonious, with respect to the sample size.

INSERT TAB. 5 HERE

The spatial dependence in the disturbances ($\lambda$) detected in both regions means that there are different unobserved factors, such as land characteristics (e.g., soil composition) and socio-economic aspects (e.g., information sharing, managerial abilities), that are spatially correlated. In other words, the coefficient $\lambda$ of the spatially autoregressive disturbances captures an *unobserved dependence structure*, so that a shock on a farm production propagates through the entire system of farms' production with a higher effect on the neighborhood. In the case of olive cultivation, for example, the closer the farm operates to the sea and the more farmers chose to grow salt- and wind-tolerant varieties. Similarly, the more the farm is located in the Northern or elevated areas, the more the farmer tends to choose frost-tolerant cultivars.

Overall, in Apulia the global model fits the data best, this means that the olive production technology does not vary over space in this region. On the contrary, in Tuscany we find the best fitting model is the one with regimes. This means that the long term, dynamic interactions among site-specific environmental variables and farmer decision making about technology have given rise to structural differences across space, hence to local technology clusters of farms. For example, Cluster 1 and 3 are very close one to the other, but the first is characterized by hills while the second by flatlands. They are among the most productive olive growing areas in Tuscany and this feature is reflected by the high and statistically significant elasticity of land. In both Cluster 1 and 3 the elasticity of capital is non-significant. In Cluster 1, this result is probably caused by the fact that the orchards are located in hilly and sloping land not suitable for mechanical operations, while in Cluster 3 it is n mainly due to the fact the trees are usually very tall and not suitable for mechanical harvesting. This latter feature is probably the cause also of the statistically significant elasticity of labor found in Cluster 3. In Cluster 2 the traditional orchards are characterized by low yields due to the habit to keep the trees low by severe pruning as a defense against frost damages. This latter feature explains the significant elasticity of labor (used for pruning and harvesting) found in this cluster. The low input management systems used in traditional orchards in this cluster explains the non-significant elasticity of other inputs and capital. In Cluster 4, the severe damage caused by harsh frosts recorded in the 1960s and 1985 led to the replacement of old trees with frost-tolerant cultivars that responded favorably to more intensive fertilizers and pesticides use. The presence of such modern plantations is probably the cause of the statistically significant response to changes in capital, i.e., mechanical work, and in other inputs, while changes in labor do not produce any statistically significant impact on output. Finally, our findings show that in all the clusters, there are decreasing returns to scale, i.e., output increases by less than the proportional change in inputs. Overall, we find that when there is spatial heterogeneity the identification of spatial regimes can help drawing a more detailed picture of the production environment and provide more accurate information to guide extension services and policy makers.

## 4. Conclusion

In agriculture, the long-term interactions among site-specific environmental variables and farmer decision making about technology have contributed to developing local specific varieties and technologies, hence to give rise to a patchy technological landscape characterized by the presence of a discrete number of spatial regimes in technologies. The clustering of farms in areas in which farms follow a local production econometric model can be the result of true contagious effects, i.e. spatial dependence, or apparent contagion effects due to structural changes, i.e. discrete spatial heterogeneity. Unfortunately, in the estimation of production functions the effects of spatial heterogeneity are often ignored. The result is that a global model is usually fitted to all the farms in the sample. Clearly, when that relationship varies over space, global parameter estimates may be very misleading and lead to wrong economic conclusions. On the other hand, it is often hard separating spatial heterogeneity from spatial dependence with only the statistical information coming from a cross-sectional data set.

In this article we use a two-step procedure that exploits the information carried by the geographic coordinates of farms to control for both spatial heterogeneity and spatial dependence. The main advantages of this procedure are that, first, it identifies spatial regimes in a data-driven approach, and, second, it estimates group-wise (regime-specific) coefficients. As a consequence, it avoids the problem of the lack of information on large sets of variables (variety grown, soil quality, etc.) causing the clustering of farms technologies in spatial regimes, and it provides more accurate information of the input-output relationships in the agricultural landscape than the global estimates..

We apply the above-mentioned two-step procedure to estimate the production function of olives on geocoded individual farm data as collected by the FADN sample survey in 2013 in 2 Italian Regions. The aim of the of this work is twofold. Firstly, we want to identify the unknown number of spatial regimes in olive growing technologies and, secondly, we want to detect the presence of spatial dependence in the data. For this latter purpose, once the spatial regimes are identified, we estimate a spatial autoregressive model with autoregressive disturbances and regimes, which captures both the spatial spillover effects (true contagion) and the heterogeneous ones (apparent contagion).

Our empirical results confirmed the existence of local technology clusters of farms, i.e. areas in which farms follow a local production econometric model. In other words, our results confirm the hypothesis that the presence of different spatial production regimes is related to the existence of a variety of latent unobserved factors, which are closely intertwined to the spatial location of the observed farms. In addition, we do not detect any statistically significant spatial autoregressive coefficient related to the dependent variables in both the regions under analysis. On the contrary, we find the spatial error dependence is at work in both regions, suggesting that the effect of a shock on a farm propagate on the other farms with the intensity given by the value of the autocorrelation coefficient.

The use of the two-step procedure to control for both spatial heterogeneity and spatial dependence can provide a detailed zoning of crop production systems that takes into account the effects of many unobserved agro-ecological and social factors at work in the territory. For example, the procedure can identify the geographical extension of specific *terroirs* and their borders, making this concept operative and relevant to land use and planning policies. In addition, by estimating local crop models with regimes we can provide regime-specific output elasticities that can be used for accurate analysis of farms' efficiencies that operate in a specific terroir.

The conditions under which the proposed partitions imply a better fit of the regression model can be further investigated. In particular, it would be interesting to extend the procedure to panel data models with both endogenous spatial regimes and spatial spillover effects in order to account for both time- and space-variations of farms production technologies.

*Econometrics: Methodology, Tools and Applications*. New York NY: Springer Verlag, pp. 99-119.

Sauer, J., and C.J. Paul. 2013. The Empirical Identification of Heterogeneous Technologies and Technical Change. *Applied Economics* 45:1461–1479.

Shee, A. and S.E. Stefanou. 2014. Endogeneity Corrected Stochastic Production Frontier and Technical Efficiency. *American Journal of Agricultural Economics*, 97(3):939–952

Stiglitz, J., and A. Atkinson. 1969. A New View of Technological Change. *Economic Journal* 79:116-131.

## Appendix

The iterative procedure to identify the spatial regimes can be summarized in the following steps.

1. Define a starting weighting vector, $w_{ij}^0 = K(d_{ij}; b^{opt})$, which is a bi-square kernel function based on both the distance between two units in space, $d_{ij}$, and the specified optimal bandwidth value ($b^{opt}$).

2. Calculate the initial pair of parameter estimates, i.e. $(\hat{\beta}_i^0, \hat{\sigma}_{\varepsilon_i}^0)$, for each location $i = 1, \ldots, n$.

3. At each iteration, say $l$, simultaneously compare the local parameter estimates $(\hat{\beta}_i^l, \hat{\beta}_j^l)$ $\forall ij$, $i \neq j$ by using Wald test statistics, $\chi_{ij}^l = (\hat{\beta}_i^l - \hat{\beta}_j^l)'(\Sigma^l)^{-1}(\hat{\beta}_i^l - \hat{\beta}_j^l)$, and calculate new weights $w_{ij}^l = K(d_{ij}^l; b)K(\chi_{ij}^l; \tau)$ as a product of the initial weights and a Gaussian kernel defined as a function of the Wald statistic values and $\tau = 0.001$.

4. To stabilize the convergence procedure, re-update the weights as $\breve{w}_{ij}^l = (1 - \eta)\breve{w}_{ij}^{l-1} + \eta w_{ij}^l$ with $\eta = 0.5$.

5. Repeat steps 2 – 4, till the condition $max|w_{ij}^{l-1} - w_{ij}^l| < \omega$ $\forall ij, i \neq j$ holds, hence till when the spatial regimes are identified, where ω is a fixed small value.

**Figures and Tables**

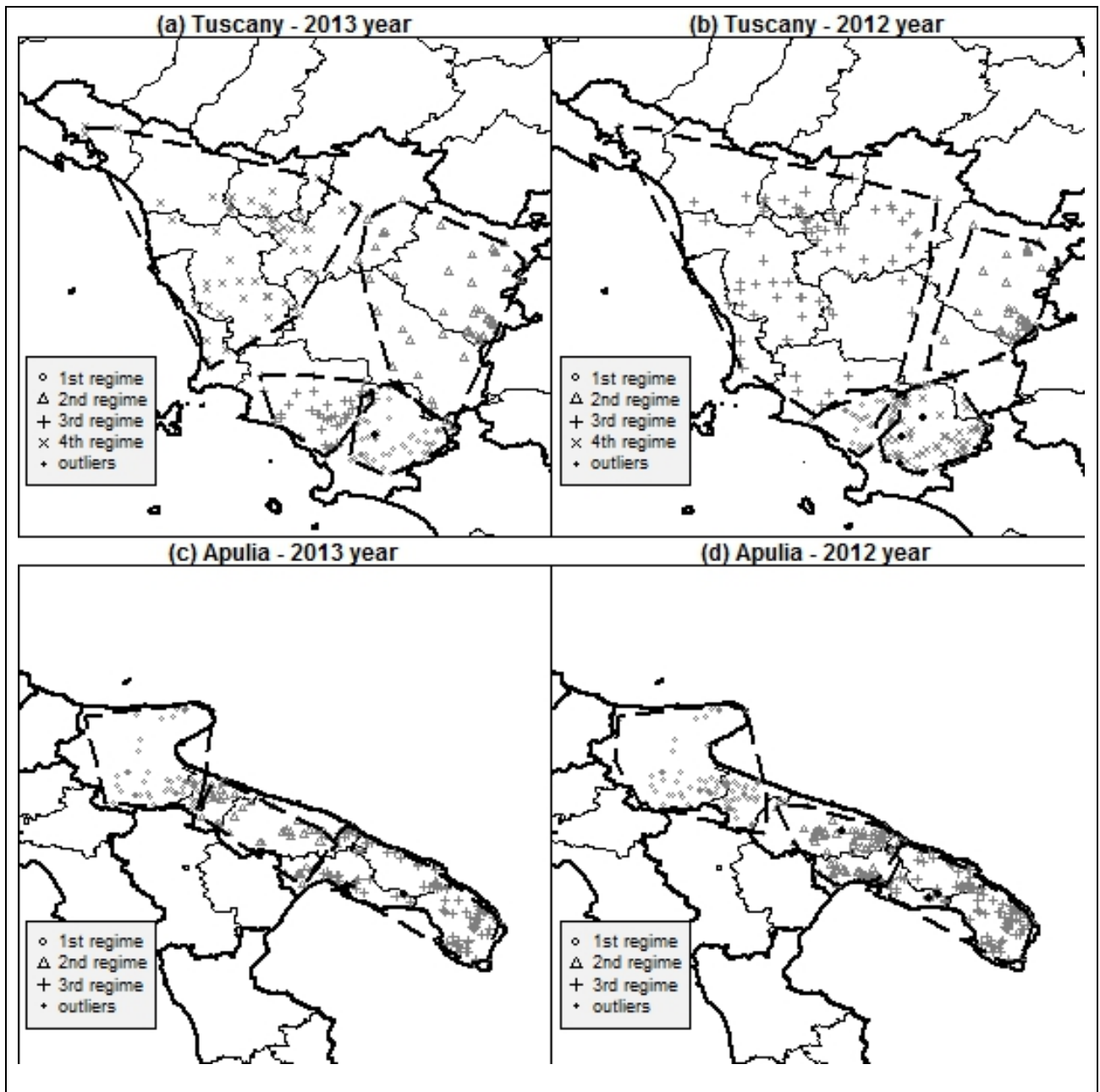

**Figure 1. Spatial technology regimes of olive farms**

*Note:* The bi-square kernel function for the initial weights and Gaussian kernel function for the updated weights are used. The adaptive bandwidths are based on the AIC criterion: (a) $b_{knn}^{AIC} = 67$, (b) $b_{knn}^{AIC} = 83$, (c) $b_{knn}^{AIC} = 111$, (d) $b_{knn}^{AIC} = 135$.

**Table 1. Descriptive statistics (year 2013)**

| Tuscany | min | median | mean | max | sd |
|---|---|---|---|---|---|
| Produced quantity | 2.000 | 31.000 | 72.420 | 1681.000 | 152.878 |
| Land (UAA) | 0.100 | 1.200 | 3.259 | 160.000 | 11.114 |
| Labor | 0.623 | 174.093 | 365.899 | 3263.510 | 489.961 |
| Capital | 0.516 | 45.000 | 92.816 | 966.4031 | 148.499 |
| Other inputs | 0.113 | 225.908 | 724.159 | 15965.754 | 1680.735 |
| **Apulia** | | | | | |
| Produced quantity | 4.000 | 130.000 | 475.400 | 8500.000 | 1008.591 |
| Land (UAA) | 0.200 | 4.100 | 12.410 | 213.000 | 28.422 |
| Labor | 73.380 | 901.830 | 1996.920 | 37894.000 | 3900.829 |
| Capital | 2.099 | 184.872 | 514.866 | 10973.854 | 1165.981 |
| Other inputs | 30.470 | 1845.760 | 3705.820 | 53250.130 | 6163.086 |

**Table 2. A-spatial and spatial Chow tests**

| | | OLS vs. OLS-regimes | | | SAE vs. SAE-regimes | | |
|---|---|---|---|---|---|---|---|
| Regions | n | t statistic | p-value | k (# par.) | t statistic | p-value | k (# par.) |
| Tuscany | 233 | 9.201 | 6.003E-08 | 5 | 31.094 | 8.975E-06 | 5 |
| Apulia | 227 | 6.127 | 2.503E-05 | 5 | 9.840 | 0.079 | 5 |

**Table 3. Spatial global models**

| | Tuscany | Apulia |
|---|---|---|
| Intercept | 1.954*** | 1.545*** |
| | (0.193) | (0.459) |
| Land (UAA) | 0.634*** | 0.733*** |
| | (0.037) | (0.061) |
| Labor | 0.099* | -0.093 |
| | (0.042) | (0.122) |
| Capital | 0.110** | 0.156** |
| | (0.040) | (0.049) |
| Other inputs | 0.096*** | 0.288*** |
| | (0.024) | (0.075) |
| lambda | 0.571*** | 0.686*** |
| | (0.088) | (0.065) |

*Note:* This table reports the global marginal effects and standard errors (in the brackets) of the log of quantity and cost variables. The units of the land are measured in hectares, whereas labor and capital are measured in working and machinery working hours, respectively. The intermediate inputs are measured in euros. Labor, capital and intermediate inputs are corrected for endogeneity. ***, **, * and . denote variables significant at the 0.1%, 1%, 5% and 10% levels, respectively.

**Table 4. Spatial autoregressive error models with regimes**

|                          | Tuscany   | Apulia    |
|--------------------------|-----------|-----------|
| Intercept Cluster 1      | 1.538**   | 0.121     |
|                          | (0.480)   | (1.000)   |
| Intercept Cluster 2      | 2.469***  | 1.975·    |
|                          | (0.266)   | (1.081)   |
| Intercept Cluster 3      | 0.641     | 1.763**   |
|                          | (0.445)   | (0.600)   |
| Intercept Cluster 4      | 3.011***  | -         |
|                          | (0.462)   |           |
| Land (UAA) Cluster 1     | 0.513***  | 0.468***  |
|                          | (0.077)   | (0.136)   |
| Land (UAA) Cluster 2     | 0.792***  | 0.884***  |
|                          | (0.062)   | (0.124)   |
| Land (UAA) Cluster 3     | 0.420***  | 0.761***  |
|                          | (0.086)   | (0.090)   |
| Land (UAA) Cluster 4     | 0.767***  | -         |
|                          | (0.070)   |           |
| Labor Cluster 1          | 0.121     | 0.139     |
|                          | (0.149)   | (0.238)   |
| Labor Cluster 2          | 0.104*    | -0.071    |
|                          | (0.052)   | (0.310)   |
| Labor Cluster 3          | 0.219*    | -0.177    |
|                          | (0.106)   | (0.163)   |
| Labor Cluster 4          | -0.159    | -         |
|                          | (0.105)   |           |
| Capital Cluster 1        | 0.110     | 0.139     |
|                          | (0.116)   | (0.086)   |
| Capital Cluster 2        | 0.065     | 0.030     |
|                          | (0.060)   | (0.144)   |
| Capital Cluster 3        | 0.162     | 0.191**   |
|                          | (0.101)   | (0.065)   |
| Capital Cluster 3        | 0.137**   | -         |
|                          | (0.070)   |           |
| Other inputs Cluster 1   | 0.148*    | 0.351*    |
|                          | (0.063)   | (0.137)   |
| Other inputs Cluster 2   | -0.006    | 0.266     |
|                          | (0.039)   | (0.214)   |
| Other inputs Cluster 3   | 0.168**   | 0.286**   |
|                          | (0.059)   | (0.097)   |
| Other inputs Cluster 4   | 0.158***  | -         |
|                          | (0.039)   |           |
| lambda                   | 0.536***  | 0.675***  |
|                          | (0.093)   | (0.066)   |

*Note:* This table reports the cluster-specific marginal effects and standard errors (in the brackets) of the log of quantity and cost variables. The units of the land are measured in hectares, whereas labor and capital are measured in working and machinery working hours, respectively. Intermediate inputs are measured in euros. Labor, capital and intermediate inputs are corrected for endogeneity. ***, **, * and . denote variables significant at the 0.1%, 1%, 5% and 10% levels, respectively.

**Table 5. AIC values for model comparisons**

| AIC | OLS | OLS-regimes | SAE | SAE-regimes |
|---|---|---|---|---|
| Tuscany | 306.953 | 293.333 | 281.215 | 280.121 |
| Apulia | 394.181 | 384.210 | 325.402 | 335.562 |

**Table 6. LR test values between SARAR and SAE specifications**

| LR test | SARAR vs. SAE | | | SARAR-regimes vs. SAE-regimes | | |
|---|---|---|---|---|---|---|
| | DF | Chisq. | Prob. | DF | Chisq. | Prob. |
| Tuscany | -1 | 1.543 | 0.214 | -1 | 0.452 | 0.501 |
| Apulia | -1 | 0.041 | 0.839 | -1 | 0.844 | 0.358 |